\documentclass[12pt]{article}
\usepackage{pst-plot,epsf}
\newcommand{\beq}{\begin{equation}}
\newcommand{\eeq}{\end{equation}}
\newcommand{\bea}{\begin{eqnarray}}
\newcommand{\eea}{\end{eqnarray}}

\newcommand{\epm}{e^+e^-}

\newcommand{\ra}{\rightarrow}

\newcommand{\eett}{e^+ e^- \ra t \bar{t}}
\newcommand{\ttbar}{t\bar{t}}
\newcommand{\bwbw}{b W^+ \bar{b} W^-}
\newcommand{\eebwbw}{e^+ e^- \ra b W^+ \bar{b} W^-}
\newcommand{\eebwsbws}{e^+ e^- \ra b {W^+}^* \bar{b} {W^-}^*}
\newcommand{\sixf}{b f_1 \bar{f'_1} \bar{b} f_2 \bar{f'_2}}
\newcommand{\eesixf}{e^+ e^- \ra b f_1 \bar{f'_1} \bar{b} f_2 \bar{f'_2}}
\newcommand{\eebnmbdu}{e^+ e^- \ra b \nu_{\mu} \mu^+ \bar{b} d \bar{u}}
\newcommand{\bnmbdu}{b \nu_{\mu} \mu^+ \bar{b} d \bar{u}}
\newcommand{\bnmbtn}{b \nu_{\mu} \mu^+ \bar{b} \tau^- \bar{\nu}_{\tau}}
\newcommand{\bcsbdu}{b c \bar{s} \bar{b} d \bar{u}}

\begin{document}
\thispagestyle{empty}
\begin{flushright}
September 2001\\
\vspace*{1.5cm}
\end{flushright}
\begin{center}
{\LARGE\bf Top quark pair production and decay into 6 fermions at
           linear colliders\footnote{Work supported
           in part by the Polish State Committee for Scientific Research
           (KBN) under contract No. 2~P03B~004~18 and by European 
           Commission's 5-th Framework contract  HPRN-CT-2000-00149.}}\\
\vspace*{2cm}
Karol Ko\l odziej\footnote{E-mail: kolodzie@us.edu.pl}\\[1cm]
{\small\it
Institute of Physics, University of Silesia\\ 
ul. Uniwersytecka 4, PL-40007 Katowice, Poland}\\
\vspace*{3.5cm}
{\bf Abstract}\\
\end{center}
The production of a $\ttbar$-pair and its decay into a 6 fermion final state
of different flavours in $\epm$ annihilation at centre of mass energies 
typical for
linear colliders is analyzed in the framework of the Standard Model. 
The results of calculation based on exact matrix 
elements at the tree level and full 6 particle phase space are 
compared with a few different approximations. It is shown that the effects
related to off-shellness of the $\ttbar$-pair and background contributions
are sizable both in the continuum and at the threshold. 
\vfill
\newpage
\section{Introduction}
The physical properties of the top quark directly measured at the Tevatron
are in a very good agreement with those derived from the Standard Model (SM)
analysis of the data collected at LEP and SLC \cite{topPDG}. However,
as the top quark is the heaviest particle ever observed, with the mass close 
to the scale of electroweak symmetry breaking, the measurement of its Yukawa 
coupling may give hints towards better understanding of the electroweak 
symmetry breaking mechanism and observed fermion mass hierarchy. 
Should the effects of the physics beyond the SM be visible at the energy scale
below 1 TeV, it is very likely that precise measurements of the top couplings 
to electroweak gauge bosons, or its electric and magnetic dipole moment 
show deviations from the corresponding SM values. The high-precision of 
measurements of the top quark properties and interactions can be
by far the best reached at an $\epm$ collider which operates at a clean
experimental environment. Therefore, such measurements are planned at TESLA 
\cite{Tesla} and will most certainly belong to the research program of any 
future $\epm$ collider \cite{NLC}.

It is clear, that in order to disentangle the possible new physical 
effects from physics of the SM, it is crucial to know the SM predictions 
for the top quark pair production and decay as precisely as possible.
Due to the large mass and width, the top quark decays before toponium
resonances can form and the predictions for 
\beq
\label{eett}
         \epm \ra t \bar{t} 
\eeq
can be obtained within the perturbative QCD. The predictions for reaction 
(\ref{eett}) in the threshold region were obtained in \cite{topQCD}
and then improved by calculation of the next-to-next-to-leading order QCD 
corrections \cite{topNNLO}, and by including the effects of initial state 
radiation and beamstrahlung \cite{topIR}.
The $\mathcal{O}(\alpha\alpha_s)$ \cite{topdec1} and 
$\mathcal{O}(\alpha\alpha_s^2)$ \cite{topdec2} corrections to the top decay 
into a $W$ boson and a $b$ quark are also known.
In the continuum above the threshold, the QCD predictions for reaction 
(\ref{eett}) are known to order $\alpha_s^2$ \cite{eettQCD} and the 
electroweak (EW) corrections to one-loop order \cite{eettEW}, 
including the hard bremsstrahlung
corrections \cite{eetthb}. The QCD and EW corrections are
large, typically of $\cal{O}$(10\%). Order $\alpha_s$ \cite{eettQCD1} 
and $\alpha_s^2$ QCD, and EW corrections has been combined in \cite{eettcomb}.

As measurements of the top quark physical characteristics, in particular its
static properties such as magnetic and electric dipole moments,
will be performed at high energies, much above the $t\bar{t}$ threshold,
it is crucial to know off-resonance background contributions to
any specific 6 fermion decay channel and to estimate the effects
related to the off-shellness of the $t\bar{t}$-pair. Therefore,
in the present note, instead of considering production of the top quark pair 
(\ref{eett}) and its subsequent decay into a specific 6 fermion final 
state, the 6 fermion reactions of the form
\beq
\label{eesixf}
         \eesixf,
\eeq
where $f_1=\nu_{\mu}, \nu_{\tau}, u, c$, $f_2=\mu^-, \tau^-, d, s$, and
$f'_1$, $f'_2$ are the corresponding weak isospin partners of $f_1$, $f_2$,
$f'_1=\mu^-, \tau^-, d, s$, $f'_2=\nu_{\mu}, \nu_{\tau}, u, c$,
are studied in the lowest order of SM.
For the sake of simplicity, it is assumed that the actual values of $f_1$ and 
$f'_2$ are 
different from each other, and that neither $f'_1$ nor $f_2$ is an electron.
The results for reaction (\ref{eesixf}) are
compared with the results obtained in a few different approximations:
the double resonance approximation for $W$ bosons
\bea
\label{doubleW}
         \eebwsbws \ra \sixf
\eea
where only those 61 Feynman diagrams are taken into account which contribute
to $\eebwbw$ and the W bosons are considered as being off-mass-shell,
the double resonance approximation for a $t$- and $\bar{t}$-quark
\bea
\label{eettsixf}
         \epm \ra t^* \bar{t}^* \ra \sixf
\eea
with only two `signal' diagrams contributing
and, finally,  with 3 different narrow width approximations:
for the $W$ bosons, top and antitop quarks,
and a single top quark \cite{BCK}.

A similar analysis of the 6 fermion processes relevant for a $\ttbar$
production in $\epm$ annihilation have been performed
in refs. \cite{ABP}, \cite{YKK}, where semileptonic channels
of reaction (\ref{eesixf}) have been studied, and
in ref. \cite{Gangemi}, where 
purely hadronic channels of (\ref{eesixf}) have been analysed.
Moreover, irreducible QCD background to top searches in semileptonic channels
of (\ref{eesixf}) has been discussed in \cite{Moretti}.
The novelty of the present work, besides the more detailed
discussion of the different approximations listed above, consists in taking 
into account both the electroweak and QCD lowest order contributions. 
Moreover, as light fermion masses are not neglected, the cross sections
are calculated without any kinematical cuts.

Basics of the calculation are described in the next section. Numerical 
results are presented and discussed in Section 3 and, finally, in Section 4,
the concluding remarks are given.

\section{Calculation}

The calculation of matrix elements of reaction (\ref{eesixf})
is based on the complete set of the Feynman diagrams at the tree level of SM. 
The number of diagrams which contribute to (\ref{eesixf}) in the unitary 
gauge, neglecting the Higgs boson coupling to fermions lighter than a
$b$ quark, amounts to 201 for semi leptonic final states, which contain
two different charged leptons, and to 333 for purely hadronic final states,
with different quark flavours.
The necessary matrix elements are calculated 
with the method proposed in \cite{KZ} and further developed in 
\cite{JK}. As in \cite{JK}, fermion masses are kept nonzero in
the matrix elements and in the kinematics. 
The constant widths of unstable particles, the massive electroweak vector 
bosons, the Higgs boson and the top quark, are introduced through the complex 
mass parameters:
\beq
\label{cmass}
M_V^2=m_V^2-im_V\Gamma_V,  \quad V=W, Z,  \qquad M_H^2=m_H^2-im_H\Gamma_H,  
\qquad M_t=m_t-i\Gamma_t/2,
\eeq
which replace masses in the corresponding propagators, both in the $s$- 
and $t$-channel Feynman diagrams,
\beq
\label{props}
\Delta_F^{\mu\nu}(q)=\frac{-g^{\mu\nu}+q^{\mu}q^{\nu}/M_V^2}{q^2-M_V^2}, \qquad
\Delta_F(q)=\frac{1}{q^2-M_H^2}, \qquad 
S_F(q)=\frac{/\!\!\!q+M_t}{q^2-M_t^2}.
\eeq
Propagators of a photon and a gluon are taken in the Feynman gauge. 

The 6 particle phase space of reaction (\ref{eesixf})
\beq
\label{dps6}
 {\rm d}^{14} Lips = (2\pi)^4 \delta^4\left( p_1+p_2-\sum_{i=3}^8 p_i\right) 
      \prod_{i=3}^8 \frac{{\rm d}^3p_i}{(2\pi)^3 2E_i},
\eeq
where the energies and momenta of the initial state particles of reaction
(\ref{eesixf}) has 
been numbered from 1 to 2, and those of the finale state particles from 3 to 8,
is parametrized in three different ways
\bea
\label{dps61}
 {\rm d}^{14} Lips  &=& 1/(2\pi)^{14}  
          {\rm d} PS_2\left(s,s_{345},s_{678}\right)
          {\rm d} PS_2\left(s_{345},m_3^2,s_{45}\right)
          {\rm d} PS_2\left(s_{678},m_6^2,s_{78}\right) \nonumber \\
&\times&  {\rm d} PS_2\left(s_{45},m_4^2,m_5^2\right)
          {\rm d} PS_2\left(s_{78},m_7^2,m_8^2\right)
          {\rm d} s_{345} {\rm d} s_{678} {\rm d} s_{45} {\rm d} s_{78},
\eea
\bea
\label{dps62}
 {\rm d}^{14} Lips  &=& 1/(2\pi)^{14}  
          {\rm d} PS_2\left(s,s_{34},s_{5678}\right)
          {\rm d} PS_2\left(s_{5678},s_{56},s_{78}\right)
          {\rm d} PS_2\left(s_{34},m_3^2,m_4^2\right) \nonumber \\
&\times&  {\rm d} PS_2\left(s_{56},m_5^2,m_6^2\right)
          {\rm d} PS_2\left(s_{78},m_7^2,m_8^2\right)
          {\rm d} s_{34} {\rm d} s_{5678} {\rm d} s_{56} {\rm d} s_{78}.
\eea
and
\bea
\label{dps63}
 {\rm d}^{14} Lips  &=& 1/(2\pi)^{14}  
          {\rm d} PS_2\left(s,m_3^2,s_{45678}\right)
          {\rm d} PS_2\left(s_{45678},s_{45},s_{678}\right)
          {\rm d} PS_2\left(s_{678},m_6^2,s_{78}\right) \nonumber \\
&\times&  {\rm d} PS_2\left(s_{45},m_4^2,m_5^2\right)
          {\rm d} PS_2\left(s_{78},m_7^2,m_8^2\right)
          {\rm d} s_{45678} {\rm d} s_{45} {\rm d} s_{678} {\rm d} s_{78}.
\eea
In Eqs.~(\ref{dps61}--\ref{dps63}), $s_{ijk...}=(p_i+p_j+p_k+...)^2,
i,j,k = 3,...,8$, and ${\rm d} PS_2\left(s,s',s''\right)$
is a two particle (subsystem) phase space element defined by
\beq
 {\rm d} PS_2\left(s,s',s''\right) = \delta^4\left( p - p' - p'' \right) 
   \frac{{\rm d}^3p'}{2E'} \frac{{\rm d}^3p''}{2E''} 
    = \frac{|\vec{p}\;'|}{4\sqrt{s}} {\rm d} \Omega',
\eeq
where $\vec{p}\;'$ is the momentum and $\Omega'$ is the solid angle
of one of the particles (subsystems) in the relative centre of mass system, 
$\vec{p}\;' + \vec{p}\;'' = 0$.
Using the rotational symmetry with respect to the beam line,
an integration over one azimuthal angle in the c.m.s. becomes trivial.
This reduces the number of necessary integrations to be performed.

Parametrization (\ref{dps61}) is most suitable for integrating the dominant
$\ttbar$ resonance contributions of Eq.~(\ref{eettsixf}).
Parametrization (\ref{dps62}) covers best contributions corresponding to
the double $W$ resonance approximation of Eq.~(\ref{doubleW}), whereas 
parametrization (\ref{dps63}) covers other  `background' contributions 
to reaction (\ref{eesixf}).
Parametrizations (\ref{dps61}--\ref{dps63}) are used with different 
permutations
of external particle momenta and with different mappings which take into 
account the Breit-Wigner shape of the $W, Z$, Higgs and top quark resonances
as well as the exchange of a massless photon or gluon.
For a given final state and c.m.s. energy, altogether about 
60 kinematical channels are sampled in order to find the dominant channels
which contribute more than 0.1\% to the total cross section. Those dominant
kinematical channels are then used in a multichannel Monte Carlo (MC)
integration routine. 

The phase space integration is simplified in the narrow width approximations.
The cross section of reaction (\ref{doubleW}) in the narrow $W$ width
approximation is given by
\beq
\label{csbwbw}
 \sigma_{\bwbw}= \sigma(\eebwbw)\;
\Gamma\left(W^+ \ra f_1 \bar{f'_1}\right)\;
\Gamma\left(W^- \ra f_2 \bar{f'_2}\right)/\Gamma_W^2.
\eeq
Similarly, the cross section of reaction (\ref{eettsixf}) in the narrow 
width approximation for the top and antitop reads
\beq
\label{cstt}
 \sigma_{\ttbar}= \sigma(\eett)\;
    \Gamma\left(t \ra b f_1 \bar{f'_1}\right)\;
         \Gamma\left(\bar{t} \ra \bar{b} f_2 \bar{f'_2}\right)/{\Gamma_t}^2.
\eeq
Finally, in the approximation where only the top quark is put on its mass
shell, the cross section is given by
\beq
\label{cst}
 \sigma_{t\bar{b} f_2 \bar{f'_2}}= 
       \sigma(\epm \ra t\bar{b}f_2 \bar{f'_2}) \;
         \Gamma\left(t \ra b f_1 \bar{f'_1}\right)/\Gamma_t.
\eeq
There are 7 integrations which have
to be performed numerically in order to obtain total cross sections 
in approximations (\ref{csbwbw}) and (\ref{cst}) and only one integration
in the case of approximation (\ref{cstt}). All numerical integrations 
in the present work are performed  with {\tt VEGAS}~\cite{vegas}. 

\section{Numerical results}
In this section, numerical results for the total and
a few differential cross sections of reaction (\ref{eesixf}) are presented.
They are compared with the corresponding results obtained within 
approximations (\ref{doubleW}), (\ref{eettsixf}) and (\ref{csbwbw}--\ref{cst}).

The SM electroweak physical parameters are defined in terms of the gauge 
boson masses and widths, the top mass and the Fermi coupling constant. 
The actual values of the parameters are taken from \cite{topPDG}:\\[4mm]
\centerline{
$m_W=80.419\; {\rm GeV}, \quad \Gamma_W=2.12\; {\rm GeV}, \qquad
m_Z=91.1882\; {\rm GeV}, \quad \Gamma_Z=2.4952\; {\rm GeV}$,}
\beq
\label{params}
m_t=174.3\; {\rm GeV}, \quad G_{\mu}=1.16639 \times 10^{-5}\;{\rm GeV}^{-2}.
\eeq
The Higgs boson mass is assumed to be $m_H=115$ GeV and the Higgs width
is calculated according to the lowest order of SM resulting in 
$\Gamma_H=4.9657$~MeV.
The top quark width is taken to be  $\Gamma_t=1.5$ GeV.

The SM electroweak coupling constants are given in terms of the electric 
charge $e_W=\left(4\pi\alpha_W\right)^{1/2}$ and electroweak mixing parameter 
$\sin^2\theta_W$ with
\beq
\label{alphaw}
\alpha_W=\sqrt{2} G_{\mu} m_W^2 \sin^2\theta_W/\pi, \qquad 
                \sin^2\theta_W=1-m_W^2/m_Z^2,
\eeq
where $m_W$ and $m_Z$ are physical masses of the $W^{\pm}$ and $Z^0$ boson
specified in Eq.~(\ref{params}). This kind of parametrization, together
with substitutions of Eq.~(\ref{cmass}), is usually referred to as the
{\em `fixed width scheme'} (FWS).
The strong coupling constant is given
by $g_s=\left(4\pi\alpha_s(M_Z)\right)^{1/2}$, with $\alpha_s(M_Z)=0.1185$.

It is also possible to perform computations with the complex electroweak
mixing parameter
\beq
\label{csw2}
\sin^2\theta_W=1-M_W^2/M_Z^2,
\eeq
with $M_W^2$ and $M_Z^2$ defined in Eq.~(\ref{cmass}). This kind of
parametrization is called the {\em `complex-mass scheme'} 
(CMS) \cite{DDRW1}. CMS has the advantage that it preserves 
the $SU(2) \times U(1)$ Ward identities \cite{DDRW1}.

For the sake of definiteness, other fermion masses used in the calculation 
are listed below \cite{topPDG}:\\[4mm]
\centerline{
$m_e=0.510998902\; {\rm MeV}, \quad m_{\mu}=105.658357\; {\rm MeV},\quad
m_{\tau}=1777.03\; {\rm MeV}$,}
\beq
m_u=5\; {\rm MeV}, \quad m_d=9\; {\rm MeV}, \quad m_s=150\; {\rm MeV}, \quad
m_c=1.3\; {\rm GeV}, \quad m_b=4.4\; {\rm GeV}.
\eeq
The Cabibo--Kobayashi--Maskawa mixing is neglected.

Matrix elements of $\eebwbw$ and $\eett$ have been checked 
against {\tt MADGRAPH} \cite{MADGRAPH} showing an agreement
up to 13--16 decimals.
As the version of {\tt MADGRAPH} used in the comparisons is not applicable
to processes with 6 particles in the final state, it has been not possible to 
compare directly matrix elements of reactions (\ref{eesixf}) 
with those generated by {\tt MADGRAPH}. 
Instead of that, matrix elements of different `subprocesses' of 
(\ref{eesixf}), namely $e^+ e^- \ra 
b f_1 \bar{f'_1} \bar{b} W^-$, $e^+ e^- \ra b W^+ \bar{b} f_2 \bar{f'_2}$
and $e^+ e^- \ra f_1 \bar{f'_1} f_2 \bar{f'_2} Z$, have been compared
successfully.
The multichannel phase space generation routine has been checked by 
comparing normalization of different channels against each other
and testing energy-momentum conservation and on-mass-shell relations.
For several total cross sections, the numerical integration has been performed 
with different parametrizations of the phase space and the results 
have been stable within one standard deviation. 

The standard deviation
of the multichannel integration routine is obtained as a sum of the
standard deviations calculated by {\tt VEGAS} for individual channels.
This gives a more conservative estimate of the integration
error than for example adding up partial errors in quadrature.

Another test is a comparison with existing calculations. Results for
total cross sections of $\eebnmbdu$ in the lowest order SM are compared with 
the results of \cite{YKK} in Table~1. As in \cite{YKK}, only the pure
electroweak diagrams and the two $\ttbar$ signal diagrams are taken into 
account in $\sigma_{\rm all \; EW}$ and $\sigma_{t^*\bar{t}^*}$, respectively.
For the sake of comparison, the physical parameters of \cite{YKK}
have been used, {\it i.e.}
$m_Z=91.187$~GeV, $\Gamma_Z=2.49$~GeV, $m_W=80.22$~GeV, $\Gamma_Z=2.052$~GeV,
$m_t=174$~GeV, $\Gamma_t=1.558$~GeV, $m_b=4.1$~GeV, $m_u=2$~MeV and 
$m_d=5$~MeV. The electroweak mixing parameter is defined as in 
Eq.~(\ref{alphaw}) and $\alpha_W=1/128.07$ is used at the same time.  
As the values of the Higgs boson mass and width used in the calculation 
are not quoted in \cite{YKK}, $m_H=115$~GeV  and the lowest order SM value 
$\Gamma_H=4.3977$~MeV, corresponding to parameters of \cite{YKK}, 
have been used in Table~1. Another source of ambiguity in the comparison
is the treatment of the finite widths of unstable particles which is not
explicitly described in \cite{YKK}. Therefore the prescription of
Eqs.~(\ref{cmass}) and (\ref{props}) have been adopted.
The results for $\sigma_{\rm all \; EW}$ and $\sigma_{t^*\bar{t}^*}$
are shown in columns 2 and 3, whereas the corresponding results of the 
present work are shown in columns 4 and 5. 
The results for $\sigma_{t^*\bar{t}^*}$ agree nicely within the uncertainties 
quoted in parenthesis. The agreement is still nice
for the complete electroweak cross sections $\sigma_{\rm all \; EW}$ above
the $\ttbar$ threshold. Below the threshold, at $\sqrt{s}=340$ GeV, there
is a substantial relative discrepancy between $\sigma_{\rm all \; EW}$ of
\cite{YKK} and that of the present work. It is amazing that the results
for all the approximated cross sections listed in Table~2 of \cite{YKK} 
agree with the present work, also at $\sqrt{s}=340$ GeV. It is difficult 
to state definitely what the actual reason for this discrepancy is. 
However, most probably it is the Higgs boson contribution, and in particular
the Higgs-strahlung `subrocess' $\epm \ra ZH$ with the Higgs boson decaying
into a virtual $W^+W^-$-pair that is responsible for it. The results
of the present work without the Higgs contribution are shown in the last 
column of Table~1. They nicely agree with
the results of \cite{YKK} which contain the Higgs with its mass and width
not being specified. 

\begin{table}
{\small Table~1: Comparison of the lowest order SM total cross sections of 
$\eebnmbdu$ of \cite{YKK} and present work. Results of \cite{YKK} obtained
with a complete set of the electroweak diagrams, $\sigma_{\rm all \; EW}$,
and two $\ttbar$ signal diagrams, $\sigma_{t^*\bar{t}^*}$, are shown in columns
2 and 3, whereas the corresponding results of the present work are shown
in columns 4 and 5. Here the parameters of \cite{YKK} are used.
All cross sections are in fb.
The number in parenthesis show the uncertainty of the last decimal.}
\begin{center}
\begin{tabular}{|c|c|c|c|c|c|}
\hline
\rule{0mm}{7mm} $\sqrt{s}$ & \multicolumn{2}{c|}{F. Yuasa, et. al. \cite{YKK}}
              & \multicolumn{3}{c|}{Present Work}\\[2mm]
\cline{2-6}
\rule{0mm}{7mm} (GeV)  & $\sigma_{\rm all \; EW}$  & $\sigma_{t^*\bar{t}^*}$
        & $\sigma_{\rm all \; EW}$ & $\sigma_{t^*\bar{t}^*}$  
                       & $\sigma_{\rm all \; EW}^{\rm no \; Higgs}$  \\[2mm]
\hline
\rule{0mm}{7mm} 
  340 & 0.687(2) & 0.4462(3) & 0769(10) & 0.4455(4) & 0.689(2) \\[1.5mm]
  350 &  6.45(1) & 6.187(4) & 6.59(2)  & 6.175(4)   & 6.45(1)  \\[1.5mm]
  360 & 14.97(2) & 14.63(1) & 15.03(5) & 14.623(9)  & 14.97(3) \\[1.5mm]
  380 & 21.42(4) & 21.00(1) & 21.48(8) & 20.99(1)   & 21.49(5) \\[1.5mm]
  500 & 22.32(4) & 21.30(1) & 22.55(4) & 21.27(1)   & 22.32(5) \\[1.5mm]
\hline
\end{tabular}
\end{center}
\end{table}

Unfortunately, a similar detailed comparison with results of 
\cite{Gangemi} is not 
possible, as the authors do not specify numerical values of the physical 
parameters used in their computations. As the cross section of reaction 
(\ref{eesixf}) at tree level is of $\cal{O}$$(\alpha_W^6)$, it is 
very sensitive to the choice of initial parameters. Although it is 
meaningless to perform
any quantitative comparison, the results of the present work are in
a qualitative agreement with those of ref. \cite{Gangemi} which will be
shown later.
A detailed quantitative comparison with Accomando, Ballestrero and 
Pizzio \cite{ABP} is also not possible as the authors include some radiative 
effects in most of their results. A meaningful comparison could in principle
be performed for the Born cross sections of $\epm \ra \bnmbdu$ 
corresponding to the $\ttbar$ signal and background at $\sqrt{s}=500$ GeV.
With cuts of \cite{ABP} and the physical parameters of the present work,
one obtains 17.895(9) fb and 1.25(2) fb for the signal and background, 
respectively.
The result for the signal cross section differs from that of \cite{ABP} 
by about 2\% while the relative difference between the background 
cross sections is much bigger, probably because there is no gluon
exchange contribution included in the Born background cross section
of \cite{ABP}. 

Lowest order SM total cross sections of the semileptonic channel $\eebnmbdu$ 
of reaction (\ref{eesixf}) at different c.m.s. energies
typical for future linear colliders are shown in Table~2. 
The complete lowest order result $\sigma$, the approximation of 
Eq.~(\ref{doubleW}) $\sigma_{b{W^+}^*\bar{b}{W^-}^*}$,
the narrow $W$ width approximation of Eq.~(\ref{csbwbw})
$\sigma_{bW^+\bar{b}W^-}$, the approximation of Eq.~(\ref{eettsixf})
$\sigma_{t^*\bar{t}^*}$, the narrow width approximation of Eq.~(\ref{cstt}) 
for a top and an
antitop quark $\sigma_{t\bar{t}}$ and the narrow width approximation
for a top quark of Eq.~(\ref{cst}) $\sigma_{t\bar{b} d \bar{u}}$
have been all obtained in FWS.
The SM tree level analytic expression for the partial widths of the 
$W$ boson and the experimental value of total $W$ width $\Gamma_W$ have been 
used in Eq. (\ref{csbwbw}).
Similarly, the SM tree level analytic expression for the partial widths of 
the $t$ quark in the zero fermion mass approximation \cite{Boos} and the 
total top width $\Gamma_t=1.5$GeV have been used in 
Eqs.~(\ref{cstt}) and (\ref{cst}).
The use of these values of $\Gamma_W$ and $\Gamma_t$ in 
Eqs.~(\ref{csbwbw}--\ref{cst}) is preferred in the comparison because the same
values have been used in substitutions of Eq. (\ref{cmass}).
In Table~2, the numbers in parenthesis are standard deviations of the 
MC integration, which show an uncertainty of the last decimal. 

A cross section of the 6 fermion reaction $\eebnmbdu$ is nonzero already
below the $\ttbar$-pair production threshold. It is the single top (antitop)
resonance
and nonresonant background contributions which are responsible for that effect.
Whether this background may affect physical observables in the threshold 
region, such as the top invariant mass distribution or angular
distributions of the final state quarks or leptons, will be discussed later.
Close to threshold, at $\sqrt{s}=360$ GeV, the 
relative difference between $\sigma$ and the narrow width approximation 
$\sigma_{t\bar{t}}$ is about --1.5\%, whereas in the continuum 
the difference becomes bigger, as relevant as radiative corrections,
amounting to 7\% at $\sqrt{s}=500$ GeV and 19\% at $\sqrt{s}=800$ GeV.
At higher energies, the difference between $\sigma$ and
$\sigma_{t\bar{t}}$ becomes so large that approximation (\ref{cstt}) does 
not make sense any more.
Comparison of approximated results $\sigma_{b{W^+}^*\bar{b}{W^-}^*}$
and $\sigma_{b{W^+}\bar{b}{W^-}}$ with the complete result $\sigma$
shows that approximations of Eqs.~(\ref{doubleW}) and (\ref{csbwbw})
are relatively much better in a wide range of the c.m.s. energy from 360 GeV
to 2 TeV.

The pure off-shellness effects of the $\ttbar$-pair can be regarded as the
difference between approximations $\sigma_{t^*\bar{t}^*}$ of 
Eq.~(\ref{eettsixf}) and $\sigma_{t\bar{t}}$ of Eq.~(\ref{cstt}).
They are plotted in Fig.~1 as a function of the c.m.s. energy.
The two plots in Fig.~1 show a similar behaviour with c.m.s. energy as those 
in Fig.~4 of F. Gangemi et. al. \cite{Gangemi}.  A naive multiplication
of the results plotted in Fig.~1 and the results for $\sigma_{t^*\bar{t}^*}$ 
of Table~2 by a factor 12, corresponding to the different
colour factor and the sum over 4 different hadronic channels,
gives nice agreement with the signal cross section
plotted in Fig.~4 of \cite{Gangemi}.

\rput(7,-6){\scalebox{0.8 0.8}{\epsfbox{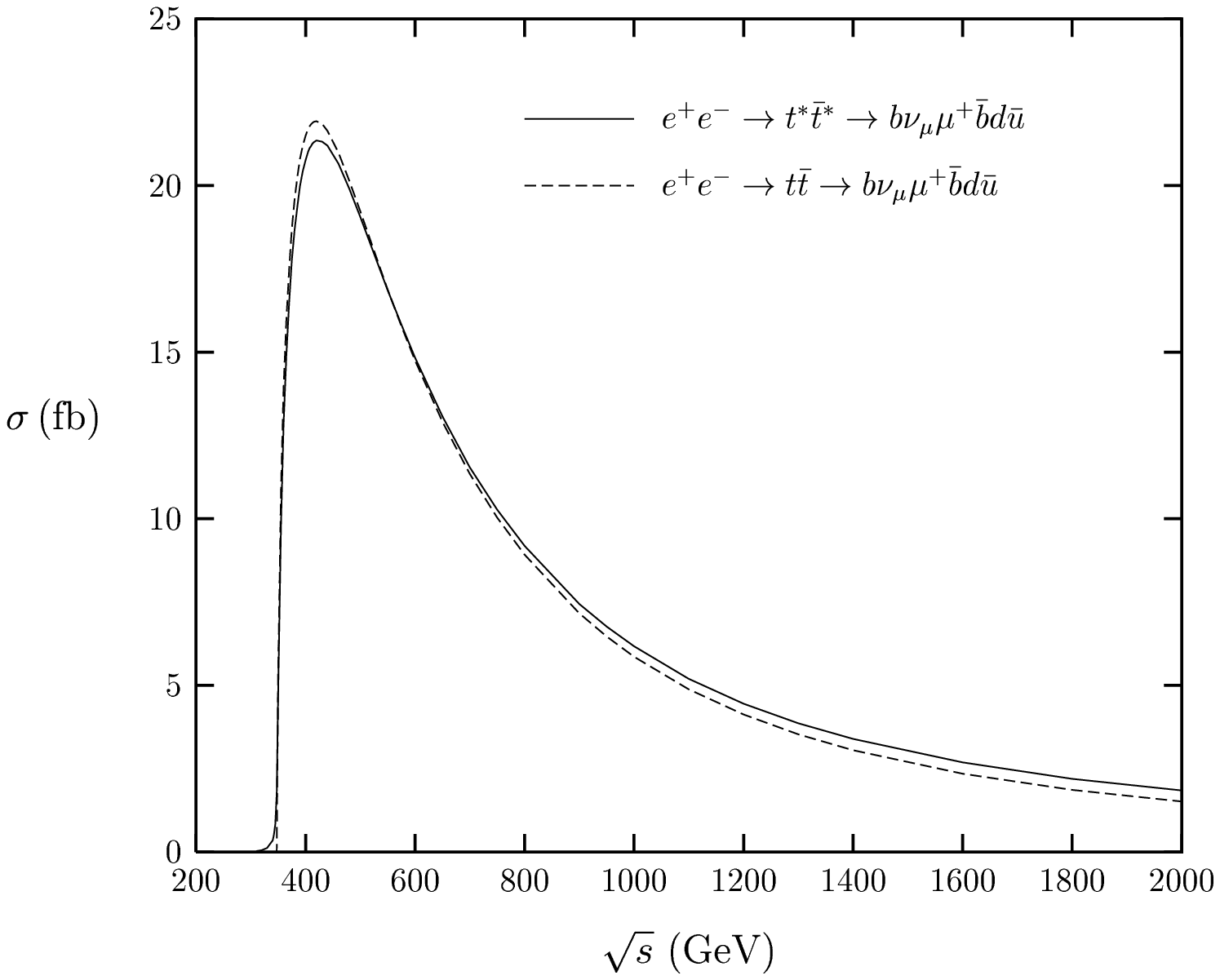}}}

\vspace*{9.0cm}

\bigskip
\bigskip
\begin{center}
{\small Figure~1. Total cross sections of $\eebnmbdu$ in approximations
of Eqs.~(\ref{eettsixf}) (solid line) and (\ref{cstt}) (dashed line)
as functions of the c.m.s. energy.}
\end{center}

\begin{table}
{\small Table~2: Lowest order SM total cross sections of $\eebnmbdu$ in fb
at different c.m.s. energies:
the complete lowest order result $\sigma$, the approximation of 
Eq.~(\ref{doubleW}) $\sigma_{b{W^+}^*\bar{b}{W^-}^*}$,
the narrow $W$ width approximation of Eq.~(\ref{csbwbw})
$\sigma_{bW^+\bar{b}W^-}$, the approximation of Eq.~(\ref{eettsixf})
$\sigma_{t^*\bar{t}^*}$, the narrow width approximation of Eq.~(\ref{cstt}) 
for a top and an
antitop quark $\sigma_{t\bar{t}}$ and the narrow width approximation
for top quark of Eq.~(\ref{cst}) $\sigma_{t\bar{b} d \bar{u}}$.
The number in parenthesis show the uncertainty of the last decimal.}
\begin{center}
\begin{tabular}{|c|c|c|c|c|c|c|}
\hline
\rule{0mm}{7mm} $\sqrt{s}$ (GeV) & $\sigma$  
& $\sigma_{b{W^+}^*\bar{b}{W^-}^*}$  
      & $\sigma_{bW^+\bar{b}W^-}$  & $\sigma_{t^*\bar{t}^*}$ 
     & $\sigma_{t\bar{t}}$ & $\sigma_{t\bar{b} d \bar{u}}$  \\[2mm]
\hline
\rule{0mm}{7mm} 
  340 & 1.162(7) & 0.681(6) & 0.671(1) & 0.3521(2) & --  & 0.2546(3) \\[1.5mm]
  360 & 13.64(2) & 13.224(8) & 13.618(8) & 12.79(1) & 13.875 & 13.42(1) 
                                       \\[1.5mm]
  500 & 20.48(9) & 20.17(1) & 20.79(1) & 19.06(1) & 19.223 & 19.51(3) \\[1.5mm]
  800 & 10.61(4) & 10.46(3) & 10.75(1) & 9.181(5) &  8.918 & 9.47(1) \\[1.5mm]
 1000 &  7.35(4) &  7.33(4) &  7.54(1) & 6.171(4) & 5.862 & 6.390(7) \\[1.5mm]
 2000 &  2.43(2) &  2.48(3) &  2.48(1) & 1.847(2) & 1.510 & 1.822(2) \\[1.5mm]
\hline
\end{tabular}
\end{center}
\end{table}

Lowest order SM total cross sections for different channels of 
(\ref{eesixf}) at c.m.s. energies typical for TESLA are compared
in Table~3. The cross sections of $\epm \ra\bcsbdu$ is about 3 times
bigger than the cross section of $\epm \ra \bnmbdu$, which in turn
is about 3 times bigger than the cross section of $\epm \ra \bnmbtn$.
This reflects the relative numbers of colour degrees of freedom. 
Small deviations of the relative factors from 3 result from the 
gluon exchange contributions, which are absent for $\epm \ra \bnmbtn$
and are different for $\epm \ra \bnmbdu$ and $\epm \ra\bcsbdu$.
The errors given in parenthesis have been obtained in the same 
way and have the same meaning as those of Tables 1 and 2.

\begin{table}
{\small Table~3: Lowest order SM total cross sections in fb for different 
top production channels at c.m.s. energies typical for TESLA.
The number in parenthesis show the uncertainty of the last decimal.}
\begin{center}
\begin{tabular}{|c|c|c|c|}
\hline
\rule{0mm}{7mm} $\sqrt{s}$ (GeV) & $\epm \ra \bnmbtn$ & $\epm \ra \bnmbdu$ & 
                                                $\epm \ra\bcsbdu$ \\[2mm]
\hline
\rule{0mm}{7mm} 360 & 4.36(1) & 13.65(4) & 42.1(2) \\[1.5mm]
                500 & 6.70(2) & 20.48(9) & 62.2(2) \\[1.5mm]
                800 & 3.43(2) & 10.61(4) & 32.1(1) \\[1.5mm]
\hline
\end{tabular}
\end{center}
\end{table}

\rput(7,-6){\scalebox{0.8 0.8}{\epsfbox{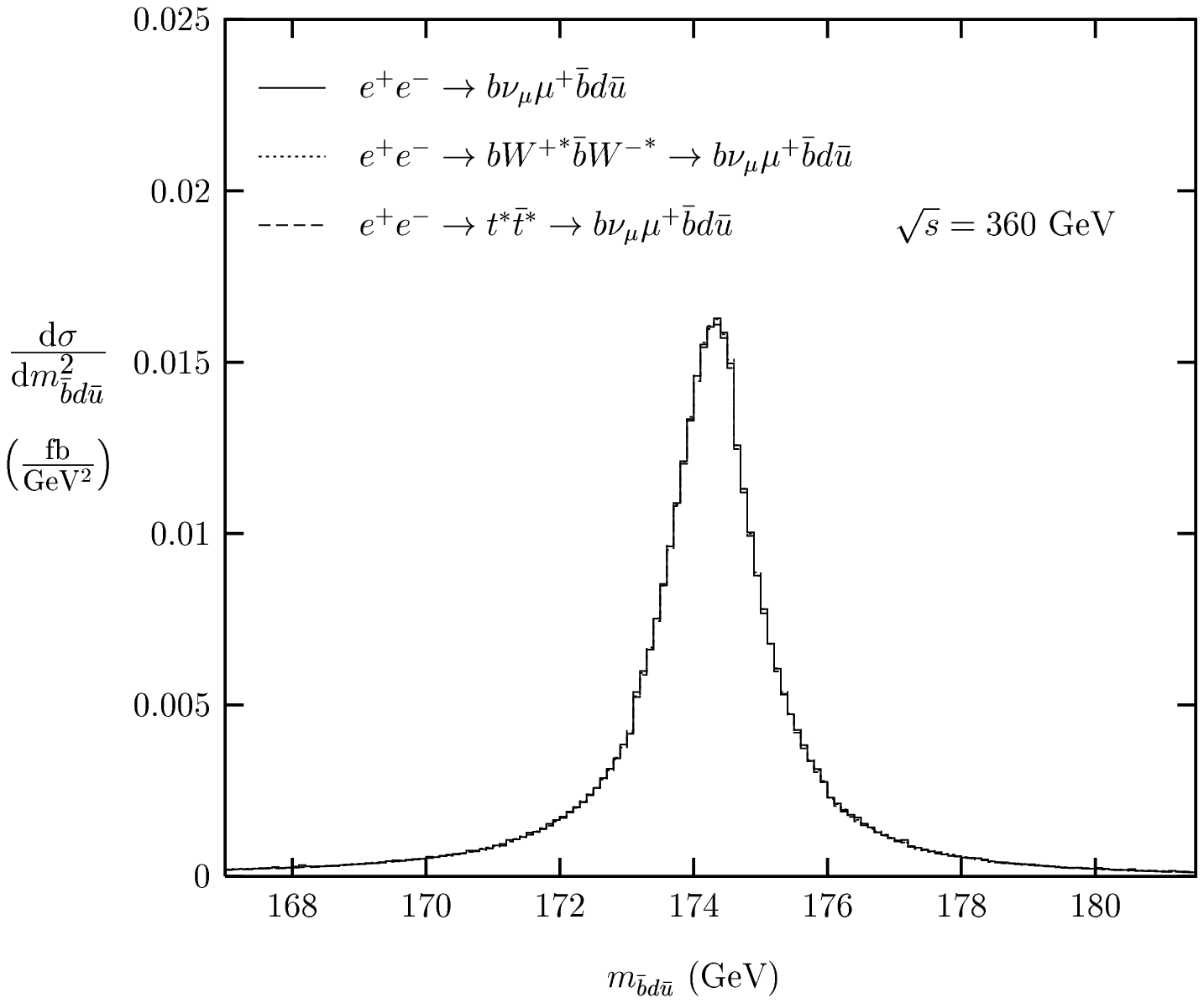}}}

\vspace*{9.0cm}

\bigskip
\bigskip
\bigskip
\begin{center}
{\small Figure~2. Differential cross sections ${\rm d} \sigma
/{\rm d} m_{\bar{t}}^2$ of $\eebnmbdu$ at 
$\sqrt{s}=360$ GeV versus the invariant mass of the $\bar{t}$ quark 
reconstructed from the $\bar{b}d\bar{u}$ system (solid histogram) and from 
the $\bar{b} W^-$ 
system in the narrow $W$ width approximation (dotted histogram).}
\end{center}

How the background nonresonant contributions affect differential cross
sections of (\ref{eesixf}) is illustrated in Figs.~2 and 3.
In Fig.~2, the differential cross sections ${\rm d} \sigma
/{\rm d} m_{\bar{b}d\bar{u}}^2$ of $\eebnmbdu$ at 
$\sqrt{s}=360$ GeV are plotted versus the invariant mass of 
the $\bar{t}$ quark reconstructed from the $\bar{b}d\bar{u}$ system.
The three histograms: solid, corresponding to the complete lowest
order result, dotted, corresponding to the approximation of 
Eq.~(\ref{doubleW}),
and dashed one, representing the $\ttbar$ signal (\ref{eettsixf}),
shown in Fig.~2, are almost indistinguishable. This means that
approximation (\ref{eettsixf}) is satisfactory and
the background contributions coming from the single top (antitop) resonance 
and the nonresonant Feynman diagrams is negligible in this case.

The differential cross sections ${\rm d}\sigma/{\rm d}\cos\theta$ 
of $\eebnmbdu$  at $\sqrt{s}=360$ GeV are plotted versus
cosine of the $\mu^+$ (up going curves) and $d$ (down going curves) 
angle with respect to the positron beam in Fig.~3. The 
angular distributions obtained with the complete set of tree
level Feynman diagrams differ substantially from the distributions based on
approximations of Eqs.~(\ref{doubleW}) and (\ref{eettsixf}).
The final state muon $\mu^+$ (down quark $d$) goes more preferably in 
the direction of initial positron (electron) than it would follow from 
the approximated distributions based on Eqs.~(\ref{doubleW}) and 
(\ref{eettsixf}).


\rput(7,-6){\scalebox{0.8 0.8}{\epsfbox{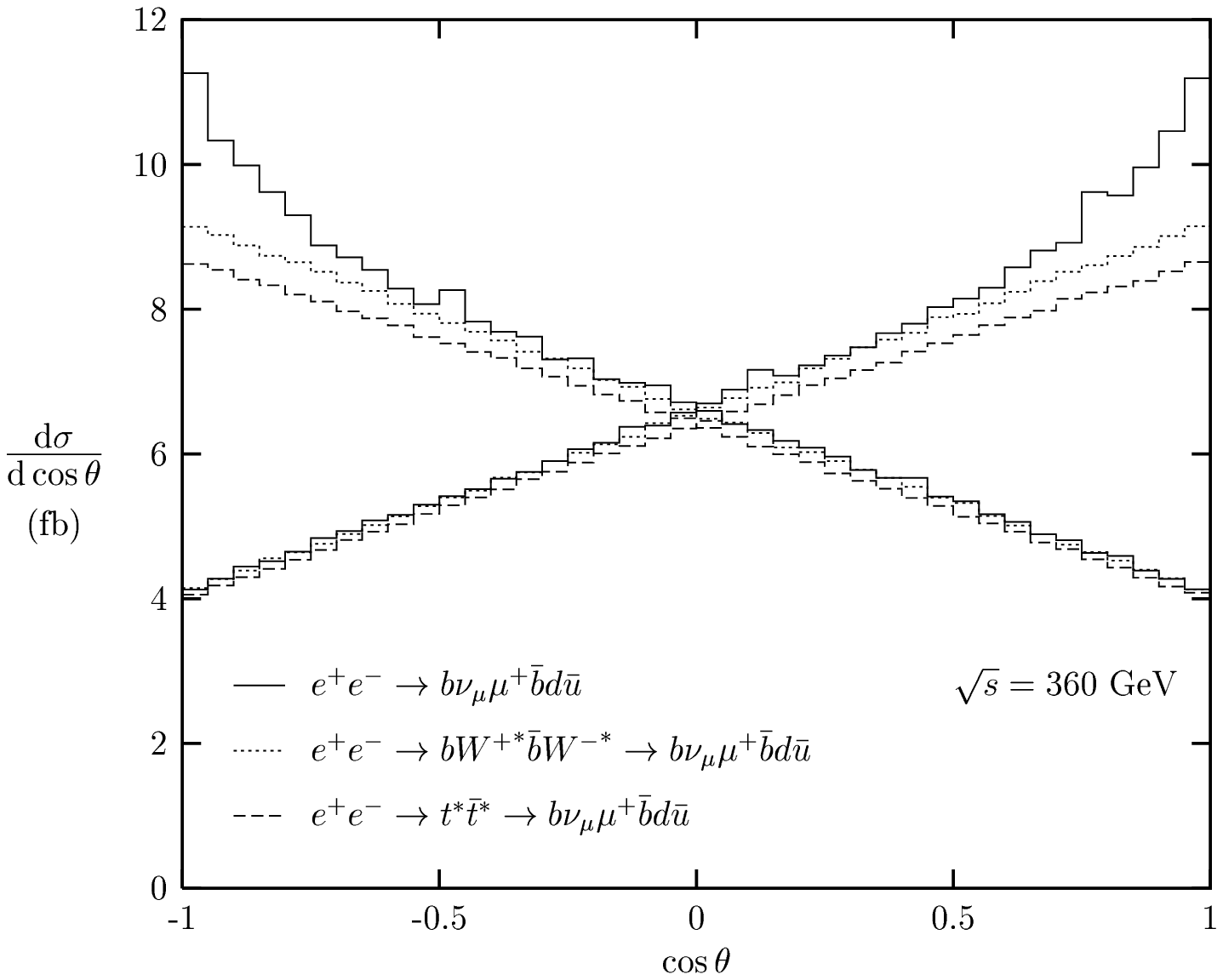}}}

\vspace*{9.0cm}

\bigskip
\bigskip
\bigskip

\begin{center}
{\small Figure~3. Differential cross sections of $\eebnmbdu$ 
at $\sqrt{s}=360$ GeV versus
cosine of the $\mu^+$ (up going curve) and $d$ (down going curve) 
angle with respect to the positron beam.}
\end{center}


\section{Summary and Outlook}

The production of a $\ttbar$-pair and its decay into a 6 fermion final state
of different flavours in $\epm$ annihilation at energies typical for
linear colliders has been analyzed in the framework of the SM. 
The results of calculation based on exact matrix 
elements at the tree level and full 6 particle phase space have been
compared with the results obtained within a few different approximations:
the double resonant approximations for the $W$ bosons (\ref{doubleW}) and for 
the top and antitop quarks (\ref{eettsixf}),
the narrow width approximation for 
the $W$ bosons (\ref{csbwbw}), the narrow width approximation for the 
$t$- and $\bar{t}$-quark (\ref{cstt}) and for the $t$-quark quark only
(\ref{cst}).

It has been shown that the effects related to the off-shellness
of the $\ttbar$-pair and to presence  of background contributions 
to cross sections of six fermion reactions (\ref{eesixf}) are quite 
substantial. They are at the level of a few per cent already in the 
$\ttbar$ threshold region. 
In the continuum, at higher energies, the effects become
quite sizable, reaching about 20\% at $\sqrt{s}=800$ GeV. Therefore,
for achieving the desired precision level in the analysis of experimental 
data from linear colliders, it is mandatory to include them 
in theoretical predictions together with radiative corrections.
The inclusion of the latter should reduce the dependence on the choice
of initial parameters mentioned in Section 3 in the context of comparisons 
with the existing calculations.


%

\end{document}